\documentclass[a4paper,8pt]{IEEEtran}
\usepackage{graphicx}    
\usepackage{cite}        
\usepackage{hyperref}    
\usepackage{url}
\usepackage[margin=1in,footskip=0.25in]{geometry}
\usepackage{subcaption} 
\usepackage{multirow} 
\usepackage{adjustbox}
\usepackage{amsmath} 
\usepackage{listings}
\usepackage{xcolor}
\usepackage{booktabs}
\usepackage{todonotes}
\colorlet{punct}{red!60!black}
\definecolor{background}{HTML}{EEEEEE}
\definecolor{delim}{RGB}{20,105,176}
\colorlet{numb}{magenta!60!black}

\lstdefinelanguage{json}{
    basicstyle=\scriptsize,
    showstringspaces=false,
    breaklines=true,
    frame=lines,
    backgroundcolor=\color{background},
    literate=
     *{0}{{{\color{numb}0}}}{1}
      {1}{{{\color{numb}1}}}{1}
      {2}{{{\color{numb}2}}}{1}
      {3}{{{\color{numb}3}}}{1}
      {4}{{{\color{numb}4}}}{1}
      {5}{{{\color{numb}5}}}{1}
      {6}{{{\color{numb}6}}}{1}
      {7}{{{\color{numb}7}}}{1}
      {8}{{{\color{numb}8}}}{1}
      {9}{{{\color{numb}9}}}{1}
      {:}{{{\color{punct}{:}}}}{1}
      {,}{{{\color{punct}{,}}}}{1}
      {\{}{{{\color{delim}{\{}}}}{1}
      {\}}{{{\color{delim}{\}}}}}{1}
      {[}{{{\color{delim}{[}}}}{1}
      {]}{{{\color{delim}{]}}}}{1},
}

\title{SD-RAG: A Prompt-Injection-Resilient Framework for Selective Disclosure in Retrieval-Augmented Generation}
\date{}

\author{
\IEEEauthorblockN{Aiman Al Masoud, Marco Arazzi, and Antonino Nocera}

\IEEEauthorblockA{Department of Electrical, Computer and Biomedical Engineering\\
University of Pavia, Pavia, Italy\\
Email: aiman.almasoud01@universitadipavia.it, marco.arazzi01@universitadipavia.it, antonino.nocera@unipv.it}
}

\begin{document}

\maketitle

\begin{abstract}
Retrieval-Augmented Generation (RAG) has attracted significant attention due to its ability to combine the generative capabilities of Large Language Models (LLMs) with knowledge obtained through efficient retrieval mechanisms over large-scale data collections. Currently, the majority of existing approaches overlook the risks associated with exposing sensitive or access-controlled information directly to the generation model. Only a few approaches propose techniques to instruct the generative model to refrain from disclosing sensitive information; however, recent studies have also demonstrated that LLMs remain vulnerable to prompt injection attacks that can override intended behavioral constraints.
For these reasons, we propose a novel approach to Selective Disclosure in Retrieval-Augmented Generation, called SD-RAG, which decouples the enforcement of security and privacy constraints from the generation process itself. Rather than relying on prompt-level safeguards, SD-RAG applies sanitization and disclosure controls during the retrieval phase, prior to augmenting the language model’s input. Moreover, we introduce a semantic mechanism to allow the ingestion of human-readable dynamic security and privacy constraints together with an optimized graph-based data model that supports fine-grained, policy-aware retrieval.
Our experimental evaluation demonstrates the superiority of SD-RAG over baseline existing approaches, achieving up to a $58\%$ improvement in the privacy score, while also showing a strong resilience to prompt injection attacks targeting the generative model.
\end{abstract}

\section{Introduction}

In recent years, Large Language Models (LLMs) are being rapidly adopted across industry and the public sector. Their use is shifting from exploratory prototypes to solutions in production environments. In industries, LLM-based solutions are adopted to ease knowledge-intensive tasks and enhance analytical workflows. By contrast, government agencies and administrations are evaluating their applicability to support a more robust digitalization and modernize administrative processes.
However, to address the limitations of LLMs, such as the lack of specific domain grounding and hallucinations, Retrieval-Augmented Generation (RAG) is becoming a favorable alternative to integrate this technology in production environments. RAG systems combine the capabilities of LLMs to elaborate and generate human language, with advanced retrieval mechanisms over curated and authoritative data corpora. Ultimately, this enables the generation of outputs that are better grounded in the specific reference domain and provide up-to-date context and specific information.
As a side effect of this trend, the private corpora of organizations are increasingly exposed to the public through the mediation of RAG pipelines. These corpora may include highly sensitive information, such as personal health records, financial data, and trade secrets.
Due to the complexity and depth of such data, it is not possible to assume a neat separation between the ``sensitive parts" and the ``safe parts", because the raw data (i.e., natural language text) are inherently unstructured. Moreover, the requirements on what is considered ``sensitive" may vary between different data owners and can change over time. 
As a matter of fact, privacy regulations may vary over time, and organization's internal privacy policies (or content moderation) can change.
To avoid disclosing sensitive information while still providing a useful level of access to information, the data owner must meticulously filter content before it is exposed to the public. However, manually checking unstructured data for sensitive details is a labor-intensive and error-prone task.
This situation motivates the need for an automated method of enforcing a human-readable privacy policy on unstructured and semi-structured text from an organization, before it is exposed to the public.
Again, modern LLMs are the natural candidate for this controlled text-manipulation task, with studies showing their usefulness in text anonymization \cite{deusser2025surveycurrenttrendsrecent} and in Redacted Contextual Question Answering (RCQA) \cite{rcqa}. 
However, the use of LLMs for content redaction presents a complication: the possibility of inducing ``prompt leaks'' by sabotaging the redaction process through prompt injection, which is an instance of the wider (and fundamental) problem of the lack of data-instruction separation in modern LLMs \cite{zverev2024can}.

Motivated by the above reasoning, in this paper we propose a novel approach for Selective Disclosure in Retrieval-Augmented Generation (SD-RAG). The proposed approach is designed to face the security and privacy challenges mentioned above and generated from the fact that existing approaches commonly expose retrieved content to the generation model. This makes RAG-based systems vulnerable to prompt injection attacks that force the model to reveal unintended or restricted information. To contrast these threats, our approach introduces a flexible mechanism for selective disclosure within RAG pipelines, enabling dynamic and fine-grained control over the portions of retrieved data that can be revealed during the subsequent LLM generation. Our approach enforces disclosure constraints as a post-retrieval processing step using the power of generative AI. This step is carried out by an internal redaction module that is in charge of enforcing security and privacy constraints.
The redacted content is hence passed to the final generation model, which only receives the minimal, sanitized data required to perform its task. 
This design choice allows the generation mechanism to remain robust even under adversarial prompting, thus balancing utility and control of unauthorized information leakage.

Another fundamental property of our proposal relies on its capability to ingest dynamic security and privacy constraints expressed in 
natural language.
To achieve this goal, we design a graph-based data model that represents the entire data corpus and encodes security and privacy information as additional nodes within the graph.
The set of nodes is hence partitioned in two subsets, namely: the data nodes and the constraint nodes.
The associations between nodes inside the graph is generated based on the semantic proximity of the concepts they represent.
Finally, in our approach, borrowing some ideas from traditional GraphRAG approaches \cite{ragWithGraphs,zhang2025survey}, we design a novel retrieval mechanism to extract not only relevant content but also associated security and privacy constraints. The constraints are retrieved by elaborating the graph relationships linking constraint nodes to data nodes. 

The main contributions of this work are hence the following:
\begin{itemize}
    \item \textit{SD-RAG}, a flexible mechanism for selective disclosure in retrieval-augmented generation pipelines that is resilient to prompt injection attacks;
    \item a dedicated graph-based data model to represent the data corpus and the security/privacy constraints;
    \item a novel mechanism to allow the ingestion of dynamic human-readable security/privacy constraints;
    \item an enhanced retrieval strategy to extract relevant content along with associated security and privacy constraints; 
    \item a generative AI–based redaction module that enforces the required constraints on the retrieved content.
\end{itemize}

In addition to the above contribution, our work also proposes a suitable strategy to evaluate SD-RAG and compare its performance with existing approaches.
Accordingly, this paper makes the following additional contributions:
\begin{itemize}
    \item an automated methodology for constructing test datasets tailored for redaction-aware closed-question answering (RCQA);
    \item novel evaluation metrics for assessing the quality of redacted content, capturing both privacy preservation and the retention of safe, useful information.
\end{itemize}

The remainder of this paper is organized as follows.
In Section \ref{sec:rw} we provide an overview of the related work and identify the research gaps therein. Section \ref{sec:methodology} outlines our methodology in terms of our assumptions, problem formulation, and proposed solution. Moreover, in this section, we introduce SD-RAG and describe its internal mechanisms. The experiments carried out along with the definition of dedicated metrics to test the quality of our solution are reported in Section \ref{sec:exp}. In Section \ref{sec:limit}, we discuss the limitations of our approach and the future research challenges and directions. Finally, in Section \ref{sec:conclusion} we draw our conclusion.

\section{Related Work}
\label{sec:rw}
Large Language Models (LLMs) demonstrate strong general capabilities, but smaller models often lack expertise in highly specialized domains. Training or fine-tuning these models requires substantial computational resources, limiting accessibility for many institutions or industries. Retrieval-Augmented Generation (RAG)~\cite{gao2024} addresses this by enriching a pre-trained LLM’s context with few-shot examples from the target domain, reducing training costs, improving modularity of indexed data, and enhancing traceability.
In basic RAG~\cite{gao2024}, document chunks are indexed as high-dimensional embeddings in a vector database and compared to the query embedding via cosine similarity. More advanced methods, such as HyDE~\cite{gao2023precise}, combine generative and contrastive techniques: a language model first generates a hypothetical document capturing the query’s intent, which is then encoded into a dense embedding for retrieval of the most relevant real documents. HyQE~\cite{zhou2024hyqe} instead generates hypothetical queries from existing contexts, measuring relevance through embedding similarity, supporting reuse of queries without fine-tuning or online generation, and complementing other ranking methods.
Tree-based indexing approaches, like RAPTOR~\cite{sarthi2024raptor}, represent both high-level and fine-grained text features, while graph-based retrieval methods (GraphRAG)~\cite{edge2024local, zhang2025survey} use a graph as an index, either solely for retrieval or also for reasoning and answer generation. More advanced RAG systems employ LLM agents that dynamically search for information, interact with external tools, and evaluate partial answers~\cite{singh2025agentic}.

While highly effective, Retrieval-Augmented Generation (RAG) systems present notable security and privacy challenges, encompassing vulnerability to prompt injection, potential data leakage, and the disclosure of sensitive information during retrieval and generation.
Recent research has highlighted several security vulnerabilities in RAG pipelines. Corpus poisoning attacks, as demonstrated by PoisonedRAG \cite{zou2025poisonedrag} and ConfusedPilot \cite{roychowdhury2024confusedpilot}, involve injecting malicious documents into the knowledge base, compromising the model’s answers and system integrity. Graph-based RAG pipelines, while more resilient, remain vulnerable to black-box attacks that manipulate inferred graph structures (GragPoison \cite{liang2025graphrag}). Additionally, RAG systems are susceptible to prompt injection, where the model’s inability to separate instructions from data can be exploited \cite{zverev2024can}.
As demonstrated by \cite{liu2023prompt}, an optimal prompt injection for many kinds of LLM-powered web applications can be automatically crafted by an LLM agent interacting with the application in a black-box setting.

Effective strategies against prompt injection have been developed, but typically require fine-tuning and can still be vulnerable to some of the more advanced white-box prompt injection techniques. \cite{chen2024struqdefendingpromptinjection}

Recent research has highlighted the risk of prompt leak in RAG systems, where sensitive information from retrieved document chunks embedded in the original prompt can unintentionally be exposed through prompt injection.

The authors of~\cite{ragDiffPriv} introduce a differential privacy redaction mechanism applied during both document retrieval and inference, without requiring LLM fine-tuning. Each indexed document is treated as a privacy unit, for instance, a single patient record, and any information overly specific to an individual is censored.

The authors of ~\cite{improvPrivBenRed}, instead, use differential privacy for model fine-tuning, combining it with Personally Identifiable Information (PII) detection, Role-Based Access Control (RBAC), and adversarial training, substantially redacting sensitive data exposure and providing protection against prompt injection attacks.

The authors of~\cite{rcqa} introduce the broader problem of Redacted Contextual Question Answering (RCQA), in which an LLM answers user queries while respecting privacy or content-moderation constraints, under the assumption that the prompter is trusted and does not attempt to inject instructions.

Although not associated to a scientific publication, \cite{protecto_secure_rag} is a relevant SaaS platform that uses RBAC to protect the context layer of LLM-powered applications, by redacting content according to some fixed sensitive data types.

A related trend in recent text anonymization research focuses on the double potential of LLMs as both: redactors (i.e. defenders) as well as de-anonymizers (i.e. attackers) \cite{deusser2025surveycurrenttrendsrecent}, with some works employing the de-anonymization capabilities of LLMs (i.e. inference attacks) to improve the process of anonymization \cite{pilan2025truthfultextsanitizationguided}, \cite{MANZANARESSALOR2025112945}.

Although automated PII detection and removal is a well-established area in machine learning, the broader and more flexible task of selective disclosure, which involves enforcing arbitrary, context-dependent constraints on what information can or cannot be revealed, remains largely unexplored, particularly in RAG pipelines, where there is a pressing need for automated redaction techniques that can operate with off-the-shelf LLMs, preserve privacy, and remain robust against untrusted, malicious, or adversarial queries.

\section{Methodology}
\label{sec:methodology}
\subsection{Threat Model}
\label{sec:threatmodel}
\begin{figure}
    \centering
    \includegraphics[width=0.5\textwidth]{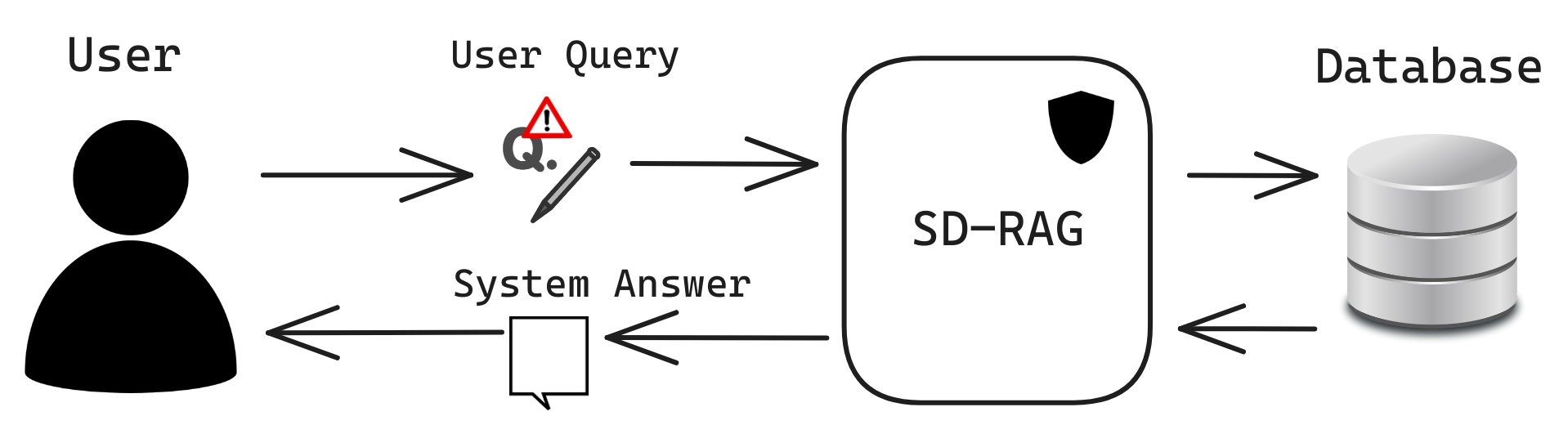}
    \caption{Threat model.}
    \label{fig:threatmodel}
\end{figure}  

Figure \ref{fig:threatmodel} illustrates our main assumptions. We assume that a user $u$ can submit arbitrary prompts $p \in P$ to the system, and each prompt is treated as untrusted input. A prompt may contain adversarially crafted instructions intended to perform prompt-injection attacks, with the goal of inducing the system to reveal the private context $C$ supplied to the language model $LM$ (for example, system instructions, hidden few-shot examples, or embedded application data). Formally, the adversary may be seen as attempting to maximize the leakage:
\begin{equation}
\max_{p \in P} \; Leakage\!\left( LM(C_r, p),\, C \right),
\end{equation}
where $C_r$ is the redacted context defined below. The adversary is assumed to have full control over the content of the prompt and seeks to maximize the likelihood that the model’s output discloses information derived from $C$.

On the other hand, the corpus indexed by the system used to build the context $C$ can be considered as trusted; this is realistic if the corpus comes from a small organization and only trusted members have write access to it. 

At each prompt $p$ from a user $u$ the context $C$, before being fed to $LM$ to generate the response, goes through a process of redaction using an additional language model $LM_r$ dedicated to apply privacy constraints on $C$ to obtain a privacy-preserving context:
\begin{equation}
C_r = LM_r(C, X).
\end{equation}
We assume that $LM_r$ is not fine-tuned on sensitive data, and thus cannot be a source of sensitive information on its own. Formally, its parameters $\theta_{LM_r}$ satisfy:
\begin{equation}
I(C;\,\theta_{LM_r}) = 0,
\end{equation}
where $I(\cdot;\cdot)$ denotes mutual information.

The constraints $X$ used to redact sensitive content from $C$ are mutually independent and do not coordinate with one another. In particular, the system does not take additional measures to ensure that a constraint that is logically entailed (either deductively or inductively) by another constraint is also enforced. That is, even if
\begin{equation}
x_i \models x_j,
\end{equation}
there is no guarantee that applying constraint $x_i$ enforces constraint $x_j$.

Finally, we assume that the database $D$ that stores the corpus with sensitive data is considered trusted.

\subsection{General Overview}

In this work, we aim to develop a versatile RAG system for selective disclosure that can adjust to evolving privacy constraints $X$, instead of focusing only on a fixed set of PII categories.
Each constraint $x_z \in X$ is a brief natural language statement that provides instructions for redacting a specific document within $C$.
Generally, a constraint refers to a sensitive category (such as "names of patients" in a medical record) and outlines the method to redact the original document to conceal the sensitive information.

Formally, a constraint may be viewed as a redaction operator:
\begin{equation}
g_{x_z} : C \rightarrow C_r^{(z)},
\end{equation}
which transforms the original context $C$ into a partially redacted version $C_r^{(z)}$.

This concept involves editing the content before presenting the context $C$ to the LLM that will respond to the user's query. This approach aims to block any harmful prompts that attempt to trick the model into ignoring the constraints $X$ when formulating its answers.

We distinguish between two redaction methods: \emph{Extractive redaction} and \emph{Periphrastic redaction}.

\textbf{Extractive redaction} utilizes the capability of an LLM to identify sensitive entities within the text that are linked to the constraints $X$ it must enforce and replace them with a placeholder. Formally,
\begin{equation}
C_r^{\text{ext}} = h_{\text{ext}}(C, X),
\end{equation}
where $h_{\text{ext}}$ performs entity detection and masking.

\textbf{Periphrastic redaction} leverages the ability of an LLM as a sentence paraphraser, instructing it to manipulate the text so that it adheres to the constraints $X$ while still retaining the general meaning of the original context $C$. Formally,
\begin{equation}
C_r^{\text{peri}} = h_{\text{peri}}(C, X),
\end{equation}
where $h_{\text{peri}}$ performs constraint-guided paraphrasing.

Finally, we formulate a ``prompt sanitization principle" that will inform our architecture: if the output of a prompt is directly returned to an untrusted actor, then the prompt must not contain both:
(i) unfiltered data from a DB (which might be sensitive), and
(ii) unfiltered user input (which might be malicious).

\subsection{SD-RAG}

\begin{figure}
    \centering
    \includegraphics[width=0.5\textwidth]{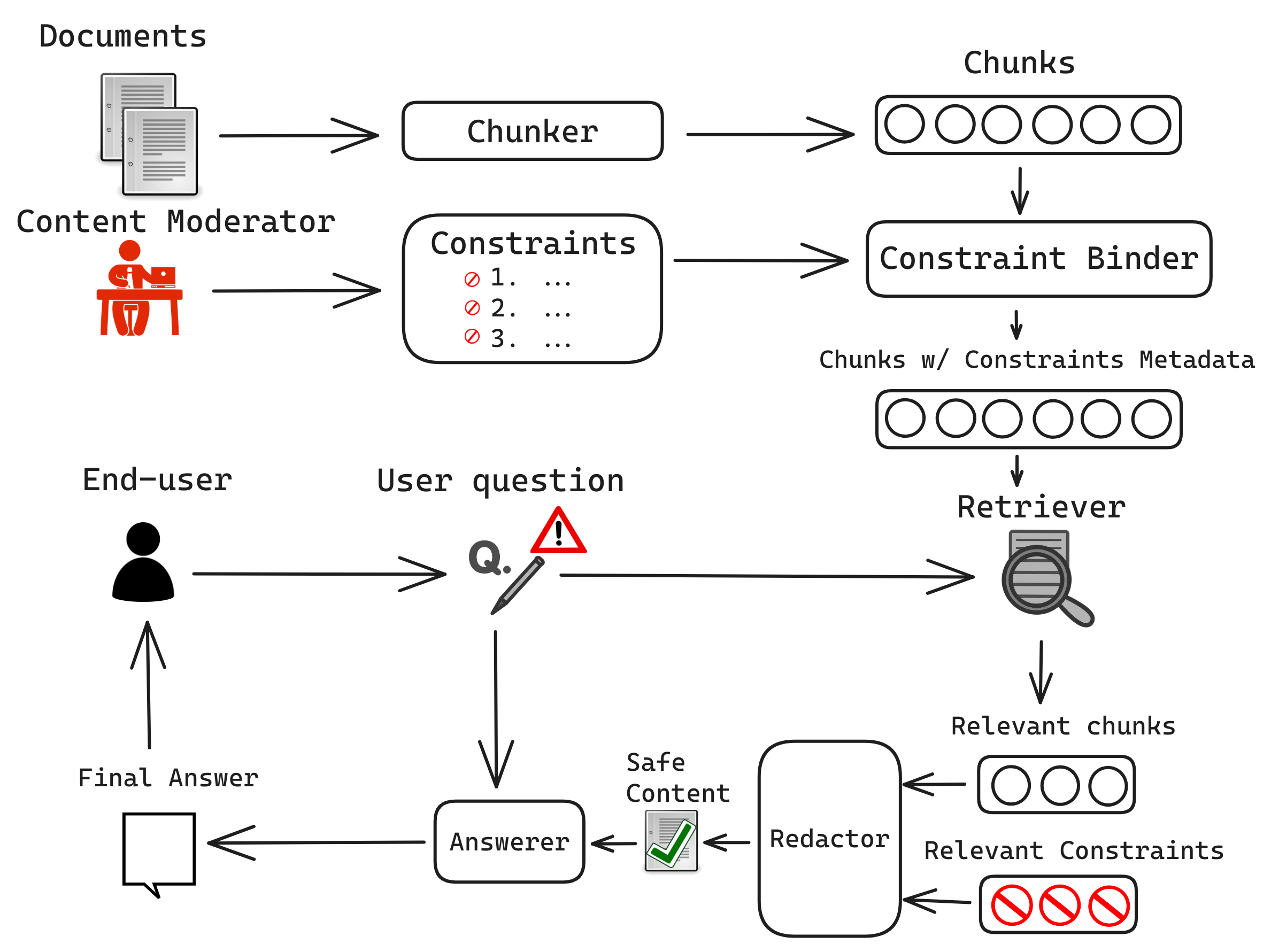}
    \caption{Architecture}
    \label{fig:arch}
\end{figure}

As we said in the previous section, we recognize that approaches that mix untrusted prompts with sensitive contexts are vulnerable to injection attacks that can bypass the privacy constraints. 
To solve this problem, we propose to let the LLM redact sensitive context separately before it is ever used to answer the user query. We allow the human redactor (or content moderator) to specify constraints as a set of high-level natural language instructions.
Figure \ref{fig:arch} offers a high-level diagram of the architecture.

The system we envision has to provide privacy enforcement on a set of documents $d_i \in D$, where the target privacy enforcement is not statically predetermined, but has to follow the custom set of constraints from $X$, introduced earlier in Section~\ref{sec:threatmodel}.

As is the case in other RAG approaches, we split each document $d_i$ into chunks $ch^{d_i}_j \in d_i$, recognizing the well-known “lost in the middle” problem of LLMs \cite{liu2023lostmiddlelanguagemodels}, where they struggle to attend to information from the central sections of long documents.

This leads us to face two distinct challenges: the first is associating the relevant constraints from $X$ to the appropriate chunks from $D$, where we assume a many-to-many relationship between chunks and relevant constraints; we will call this: the \emph{constraint binding} problem.

The second challenge is to actually redact the chunks once the constraints that apply to them are known; we will call this: the \emph{ constraint application} problem.

\label{sec:constraint_binding}
\subsubsection{Constraint Binding Mechanism}

the problem of binding the constraints to the chunks they apply to can be effectively seen as a RAG within the RAG. 

\begin{figure}
     \centering
     \includegraphics[width=0.8\linewidth]{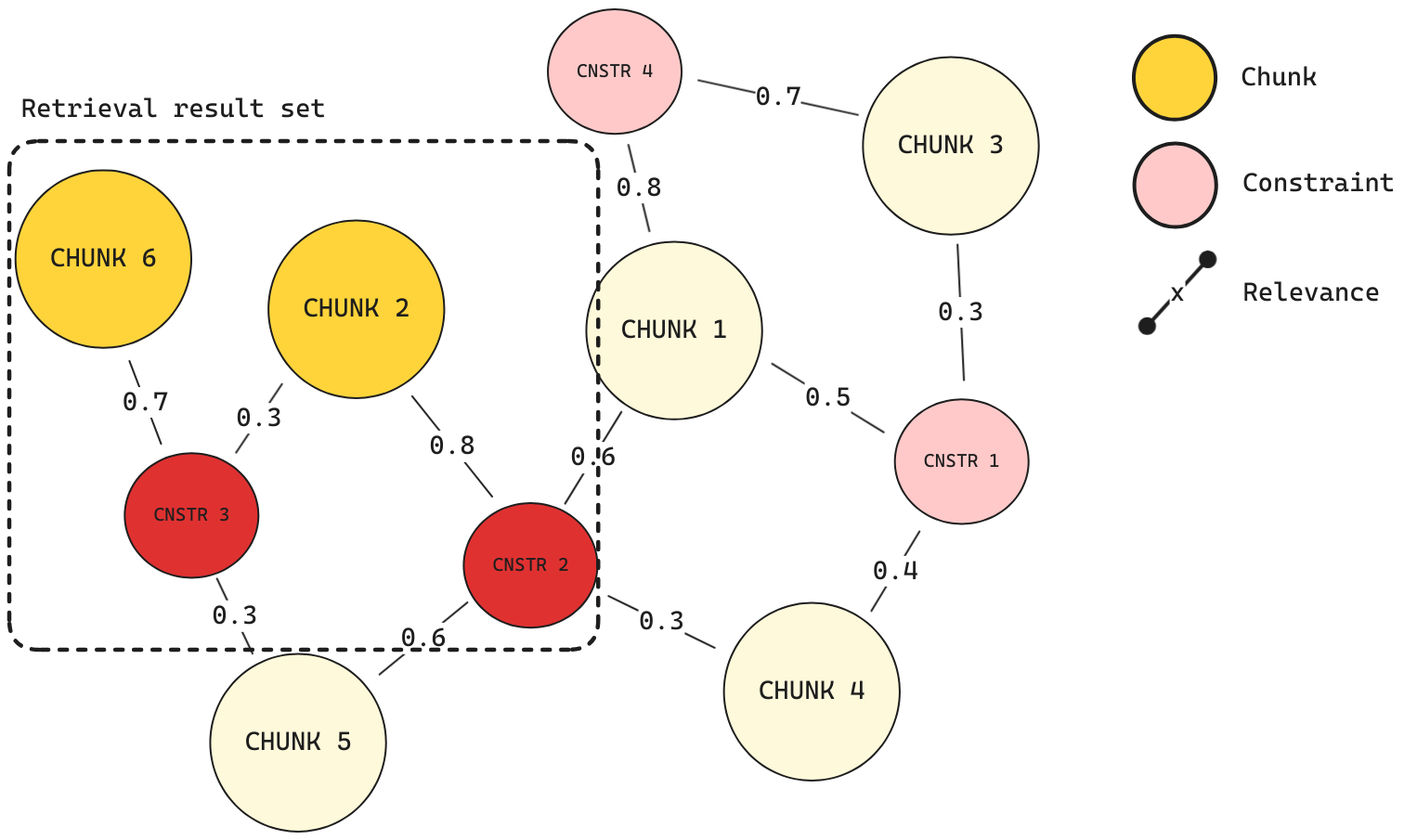}
     \caption{Constraint-to-chunk binding}
     \label{fig:constraint-attachment}
 \end{figure}

We begin at indexing time, after splitting the documents $D$ into chunks:

\begin{enumerate}
    \item {the constraint binder attaches each constraint to the top $N$ most similar chunks, retrieved by cosine similarity of text embeddings;}
    \item {as illustrated in figure \ref{fig:constraint-attachment} each chunk stores its similarity score to each of the attached constraints; this is useful for some constraint re-ranking strategies that we introduce in Section \ref{sec:constraint_reranking}.}
\end{enumerate}

This constraint binding strategy is linear in the number of constraints, as $N$ is a fixed parameter that the system operator can choose.

At retrieval time, after the user query was used to retrieve the set of chunks $C$:

\begin{enumerate}
    \item {For each chunk $c_i \in C$, we retrieve the set of constraints $x_i \in X$ that are bound to $c_i$ from indexing time.}
    \item {We consider the union of these sets as the initial candidate constraints.}
    \item {We rank these candidate constraints according to a re-ranking strategy (as detailed in section \ref{sec:constraint_reranking}).}
    \item {We only keep the top $K$ constraints, and apply them to the chunks as detailed in section \ref{sec:constraint_application}.}
\end{enumerate}

We apply the constraints to the chunks only at retrieval time, not an indexing time.

We do this because: (i) redaction of chunks is an expensive operation that does not need to be paid upfront, for example, an attached constraint may never be used if it is removed following a change in privacy policy; and (ii) some constraints are more relational than others, and can be activated by extra context that is not present in a single chunk.

To confer more flexibility to our architecture, we introduce an optional hierarchical chunk summarization module inspired by \cite{sarthi2024raptor}.

This enables us to examine whether a hierarchical summarization scheme provides any advantages for privacy protection. We hypothesize that hierarchical summarization may enhance privacy through two mechanisms:

\begin{itemize}
  \item Summarization involves generalization of specific (and potentially private) details.
  \item Summarization may bring together multiple chunks that would not individually trigger a redaction constraint but that collectively do.
\end{itemize}

As per \cite{sarthi2024raptor}, the hierarchical summarization module computes the embeddings of the chunks, projects them to a lower dimension vector space using UMAP \cite{mcinnes2020umapuniformmanifoldapproximation}, and clusters the reduced-dimensionality embeddings using a Gaussian Mixture Model (GMM).

Each cluster of chunks is then converted into a summary by an LLM, and the whole process is repeated on the summaries to create higher-order summaries, which are linked to the nodes that sourced them. This process is iteratively repeated until a certain tree depth (i.e. number of layers) is reached.

The nodes of the resulting tree capture information at various levels of granularity. We then ``flatten" the tree and introduce its nodes as chunks into the index.

\label{sec:constraint_application}
\subsubsection{Constraint Application}

The actual redaction that applies the constraints $X^{ch^{d_i}_j}$ on the relative chunks $ch^{d_i}_j$ is performed by the $LLM$ using a safe prompt that does not include any query from the user, producing the summaries just before answering the user query.

Specifically, after retrieving the top document $d_i$ chunks $CH^{d_i}$ up to a predetermined token budget, they are concatenated into a single string. Formally, let
\begin{equation}
\mathrm{Context}(d_i) \;=\; \mathrm{concat}\!\left( CH^{d_i} \right).
\end{equation}

This resulting context string is redacted using the $K$ retrieved constraints $X^{d_i}$ (as per section \ref{sec:constraint_binding}).

Both periphrastic and extractive redaction are implemented as few-shot prompts that apply a batch of constraints $B \subset X^{d_i}$ at a time to the context.

To perform periphrastic redaction on a text, we iteratively rephrase it according to each batch of constraints, and reassign it after each iteration.

To perform extractive redaction on a text, we iteratively extract the spans of text determined by each batch of constraints, but we do not reassign the text: we always perform the extraction on the original, and substitute the identified sensitive spans all at once only at the end of the process. We find that this method is less error prone, because it does not introduce ``placeholder" strings at each iteration.

The redacted context is then used to answer the user query. The confidentiality impact of a potential prompt injection in the query is thus minimized, assuming that the chunks themselves are benign and the redaction method is effective.

\section{Experiments}
\label{sec:exp}

This section is devoted to present the experiments conducted to validate the approach.

\subsection{Experimental Setup}
\label{sec:expsetup}

\subsubsection{Dataset}
\label{sec:dataset}

Because no existing dataset exactly matches our scenario, we began by crafting one ourselves, which we propose as an additional contribution of this paper.

The proposed dataset consists of short synthetic articles about various topics, each containing some examples of "semi-sensitive" information of different categories.

Each article is associated to a set of Q\&A pairs, and each  Q\&A pair contains a question that pertains to some of the sensitive categories of information, but can be profitably answered without disclosing any, as demonstrated by the golden answer.

This is an example of how an annotated article from the test set can look like:

\begin{lstlisting}[language=json,firstnumber=1]
// This is one article:
{
    "category": "criminal",
    "article_text": "In the early hours of March 12th...",
    "constraint_to_witness_words" : {
        "Remove names of witnesses": ["Jonathan", "Meyers", /* ... */ ],
        // More constraints...
    },
    "questions" : [
        {
            "question_text": "How did witnesses first notice the suspects?",
            "answer_text": "One of the witnesses...",
            "relevant_constraints": [
                "Remove names of witnesses",
                "Remove names of perpetrators"
            ]
        }
        // More questions...
    ]
}

\end{lstlisting}

We adopt a semi-automated strategy to build the proposed dataset, comprising the following steps:

\begin{enumerate}

    \item{We prompt Gemini 2.5 Flash \cite{comanici2025gemini25pushingfrontier} to generate a list of topics for an article that may contain sensitive data, together with a list of likely sensitive categories of information for each topic. For instance, an article topic may be "a medical record" and the sensitive information may be: "names of patients" or "specific doses of medicine"}
  
    \item{We prompt the LLM to actually write a short synthetic article for each topic, instructing it to include instances of all the sensitive categories of information. In addition to generating the article, we also instruct the LLM to return a mapping from each sensitive category of information to its specific instances in the article.}
  
    \item{We prompt the LLM to generate a list of Q\&A pairs that can be asked about and answered by the article.}
  
    \item{We associate each Q\&A pair to its set of relevant sensitive categories, by iterating over each category and checking if any of its instances is included in the answer or in the question of the Q\&A pair. We do this using a keyword-based criterion.}
  
    \item{We prompt the LLM to remove any explicit reference to the instances of the sensitive categories (which we have from step 2) from both the answers as well as the questions.}
  
    \item{We manually inspect the generated data for consistency and correctness.}
  
\end{enumerate}

\subsubsection{Metrics}

To evaluate the performance of the system, we introduce two metrics to measure the quality of redacted answers: (i) Privacy Score and (ii) Completeness score. The former evaluates how effectively the generated answer prevents the disclosure of sensitive information, and the latter measures how informative (in a safe way) the answer still manages to be.

\paragraph{Privacy Score}

The privacy score of an answer represents the percentage of relevant constraints that the answer satisfies.

\begin{equation}
P(A) = \frac{ | C_{R}(A) \cap C_{S}(A) | }{ | C_{R}(A) | }\end{equation}

Where: 

\begin{itemize}
  \item{ $P()$ is the privacy score, }
  \item{ $A$ is the answer to a question, }
  \item{ $C_{S}(A)$ is the set of satisfied constraints of answer $A$, }
  \item{ $C_{R}(A)$ is the set of relevant constraints of answer $A$.}
\end{itemize}

\textbf{Example}
Consider the question: ``Which cardiologist did the patient consult, and where?'' 
Suppose that the two relevant privacy constraints for this query are:
(i) hide the patient's name ($x_{\text{patient-name}}$), and 
(ii) hide the doctor's name ($x_{\text{doctor-name}}$).

Assume that the model produces the answer: 
``The patient consulted Dr.\ Smith at the Springfield Heart Institute.''

We then evaluate whether each applicable constraint is satisfied. In this case, the answer
does not reveal the patient's name, but it does reveal the doctor's name. Hence, out of the two
relevant constraints, only one is satisfied, yielding a privacy score of
\begin{equation}
P = \frac{1}{2} = 0.5. \notag
\end{equation}

\label{completenessscore}
\paragraph{Completeness Score}

The completeness score of an answer (between 0 and 1) measures the percentage of safe target facts that are included in the answer. 

\begin{equation}
\small
C(A) = \frac{\sum_{f_G \in F(A_G)}{ max \{CosSim(f_G, f_A) : f_A \in F(A) \}}}{| F(A_G) |}
\end{equation}

\begin{itemize}
    \item{$C()$ is the completeness score,}
    \item{$F()$ is a function that returns the facts/claims implied by an answer,}
    \item{$F(A)$ is the set of claims of answer $A$,}
    \item{$F(A_G)$ is the set of claims of {\bf golden} answer $A_G$,}
    \item{$max \{ sim(f_G, f_A) : f_A \in F(A) \}$ is the score of the fact of $A$ (i.e. $f_A$)  most similar to the golden fact $f_G$.}
\end{itemize}
  
\textbf{Example.}
Consider the question: ``What medications were prescribed to the patient as part of his treatment plan?'' 
Suppose the golden answer states: 
``He was prescribed a medicine to manage his blood pressure, and another medicine to prevent clot formation.''

Now, assume the model outputs: 
``The patient was prescribed a medicine to manage his blood pressure.''

To evaluate completeness, we compare the answer against the set of target facts. In this case, the answer includes only one of the two target facts. Therefore, the completeness score is
\begin{equation}
C = \frac{1}{2} = 0.5, \notag
\end{equation}
since it misses the second fact: the patient was also prescribed a ``medicine to prevent clot formation.''

\subsection{Metrics Operationalization}

\label{sec:privscoreop}
\subsubsection{Privacy Score}

We operationalize this metric using the concept of ``witness words" borrowed from \cite{zverev2024can}. Each test question has a set of associated ``relevant constraints" and each of these constraints has a set of ``witness words"; if a witness word appears in an answer, we can reasonably be sure that the constraint was broken. For instance: the witness word ``Dr. Smith" if the constraint says to "hide names of doctors".

\subsubsection{Completeness Score}

We implement a completeness score based on the original formula detailed above \ref{completenessscore},using a generative LLM to extract facts from the golden and actual answers, and a text embedding model to compute similarity scores between the extracted facts.

\subsection{Results}

\subsubsection{Robustness against Prompt Injection}

The aim of this section is to evaluate the performance of our approach against prompt injection attacks and, simultaneously, to compare it with the state-of-the-art baseline introduced in \cite{rcqa}. The baseline approach relies on a “monolithic prompt” that embeds the constraints directly into the prompt to produce the final answer, while also incorporating the user’s potentially malicious query.
The comparison is performed in terms of privacy and completeness scores introduced in Section~\ref{sec:privscoreop}.
For this experiment, we treated each of articles that compose our synthetic dataset (Section~\ref{sec:dataset}) as a separate corpus, building an index for each, and running the Q\&As with each defense strategy.

For prompt injection, we manually craft a simple template that instructs the LLM to disregard its previous instructions (whatever they may be) and to instead output the full context it was given by the previous prompts.

We run each experiment on two open-source LLMs: Qwen2.5 7B and Llama-3 8B, both quantized to q4\_k\_m quality. 

We always use the bigger model (i.e. Llama-3 8B) to evaluate the completeness score. 
We compute the average privacy score and completeness score, aggregating by defense strategy and by presence of prompt injection, which is shown in Table \ref{tab:injection}.

\begin{table}[ht]
\centering
\resizebox{\columnwidth}{!}{
\begin{tabular}{llrr}
\toprule
\textbf{Defense} & \textbf{Injection} & \textbf{Privacy Score} & \textbf{Completeness Score} \\
\midrule

\multirow{2}{*}{Baseline~\cite{rcqa}} 
  & False & 0.502438 & 0.689494 \\
  & True  & 0.198961 & 0.583328 \\
\midrule

\multirow{2}{*}{Rephrase} 
  & False & 0.728657 & 0.621787 \\
  & True  & 0.585128 & 0.583601 \\
\midrule

\multirow{2}{*}{Extract} 
  & False & 0.826739 & 0.597963 \\
  & True  & 0.778937 & 0.582014 \\
\bottomrule
\end{tabular}
} 
\caption{Privacy and completeness scores for different defenses with and without prompt injection.}
\label{tab:injection}
\end{table}

Considering a benign user query (i.e. no prompt injection), we can see that our method (``rephrase" and ``extract") already perform better in terms of privacy score than the baseline; indicating that even when the model is not attacked and is still following the original Redacted Contextual Question Answering (RCQA) monolithic prompt it will still tend to leak sensitive information. 

Our method reduces up to 30\% of that information leak by removing the sensitive information before it is ever used to produce the final response, with extractive redaction generally achieving a higher privacy score.

Conversely, the monolithic prompt achieves a slightly higher completeness score (up to 10\%), suggesting that the pre-redaction step also discards some useful non-sensitive content. In other words, extractive redaction appears to operate more aggressively, sacrificing some completeness in exchange for its higher privacy score.
Under a malicious query (i.e. with prompt injection), the LLM generally abandons its original task and, instead, leaks the full context it was given. In the baseline strategy, the monolithic prompt receives unredacted chunks with sensitive information; thus, when the monolithic prompt is ignored as a result of the injection, the original sensitive chunks all end up in the final response, causing a major drop in privacy score.

With both of our pre-redaction strategies, the effects of a prompt injection are reduced, since the chunks are redacted independently of the final question-answering prompt. 
We notice how Extractive redaction experiences a lower drop in privacy score, and still outperforms both the baseline and periphrastic strategies in the scenario of no attack.

When the attack is applied, the completeness score falls to approximately the same level across all strategies. This suggests that the LLM behaves almost identically in every case: it abandons its initial task and simply reproduces the entire context it received.  
The target answers (used to compute the completeness score) are constructed to contain as little sensitive information as possible, and the full context (whether redacted or not) is less informative than the specific answer we expect. As a result, every strategy under attack yields a similar completeness score.

\label{sec:constraint_reranking}
\subsubsection{Effect of Re-ranking on Constraint Retrieval}

as detailed in section \ref{sec:constraint_binding}, we bind the constraints to the chunks at indexing time, and then we employ a two-stage strategy to retrieve the constraints most appropriate to the relevant chunks. 
The first step involves finding all candidate constraints, i.e., all of the ones attached to the relevant chunks.
The second step involves ranking the candidates and selecting only the top K matching ones.

In this experiment, we compare the performance of several constraint re-ranking techniques (all based on dense embeddings) in terms of recall at K (R@K) and average precision at K (AP@K).
To compare the performances of the re-ranking methods, we built a single corpus from the full 20 articles in our synthetic dataset (totaling at about 13600 tokens) which were split into 639 chunks. 
There were a total of 120 constraints, and each constraint was initially attached to the 50 most similar chunks.
We have focused on recall, because we believe that in our scenario minimizing false negatives (privacy constraints that should be applied but are not) is more important than avoiding false positives (constraints that should not be applied but are applied). 

We compared the re-ranking methods described in details below.

\begin{figure}[h]
    \centering
    \includegraphics[width=0.9\linewidth]{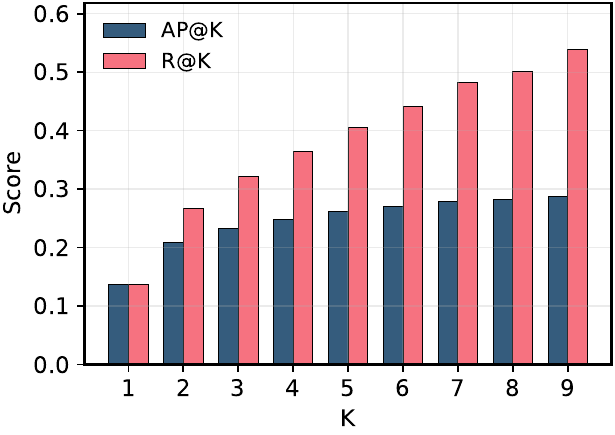}
    \caption{Re-ranking constraints by similarity to concatenated chunks}
    \label{fig:rerank_context}
\end{figure}

\textbf{Re-ranking constraints by similarity to concatenated chunks}: $score_{cat}(c) = sim(c, concat(D))$ this method ranks the candidate constraints based on their similarity (i.e. cosine similarity of dense embeddings) to the concatenation of the retrieved chunks (i.e. the relevant context). The intuition is that a constraint's relevance must be evaluated in the holistic context that will be used to answer the query The results are shown in Figure \ref{fig:rerank_context}.

\begin{figure}[h]
    \centering
    \includegraphics[width=0.9\linewidth]{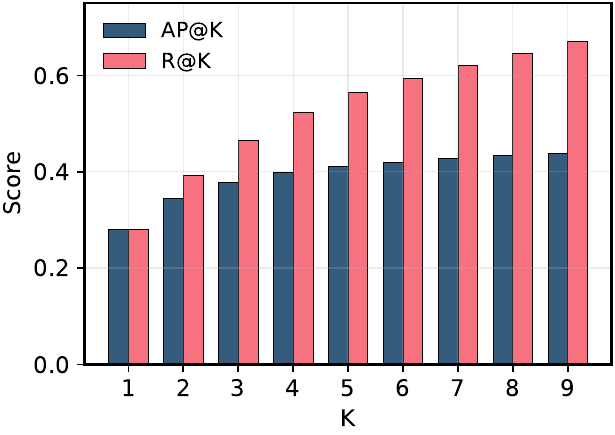}
    \caption{Re-ranking constraints by similarity to query}
    \label{fig:rerank_query}
\end{figure}

\textbf{Re-ranking constraints by similarity to query} $score_{q}(c) = sim(c, query)$ This method ranks the candidate constraints based on their similarity to the user query that was used to retrieve the constraints. The intuition is that the precise importance of each attached constraint is in part determined by the use that will be made of the chunk, which is determined by the query. The results in Figure \ref{fig:rerank_query} are overall a bit better than using the concatenated chunks.

\begin{figure}[h]
    \centering
    \includegraphics[width=0.9\linewidth]{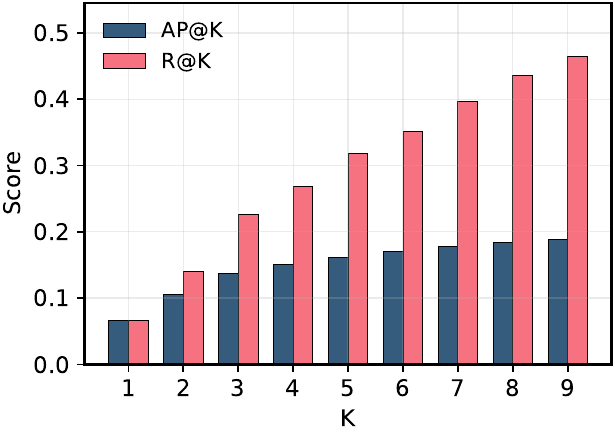}
    \caption{Re-ranking constraints by maximum similarity to retrieved chunks}
    \label{fig:rerank_max}
\end{figure}

\textbf{Re-ranking constraints by maximum similarity to retrieved chunks} $score_{max}(c) = max (sim(c, d)) | d \in D$ This method ranks each candidate constraint based on its maximum similarity to any of the retrieved chunks. The results in Figure \ref{fig:rerank_max} are worse than the previous two methods;  we think that this is due to some constraints having very high relevance score to one chunk, but this chunk not being relevant to the query, rendering the constraint also irrelevant.

\begin{figure}[h]
    \centering
    \includegraphics[width=0.9\linewidth]{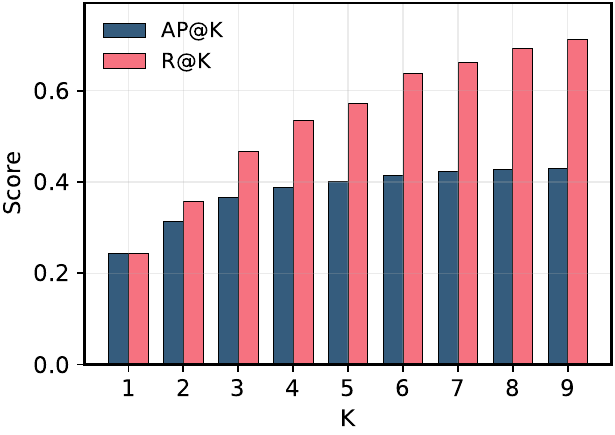}
    \caption{Re-ranking constraints by average similarity to retrieved chunks}
    \label{fig:rerank_avg}
\end{figure}

\textbf{Re-ranking constraints by average similarity to retrieved chunks} $score_{avg}(c) = \frac{1}{D} \sum_j^D sim(c, d_j)$ This method ranks each candidate constraint based on its average similarity to the retrieved chunks, and the results in Figure \ref{fig:rerank_avg} are better than the previous method. We conclude that averaging helps reduce the influence of "outlier constraints".

\begin{figure}[h]
    \centering
    \includegraphics[width=0.9\linewidth]{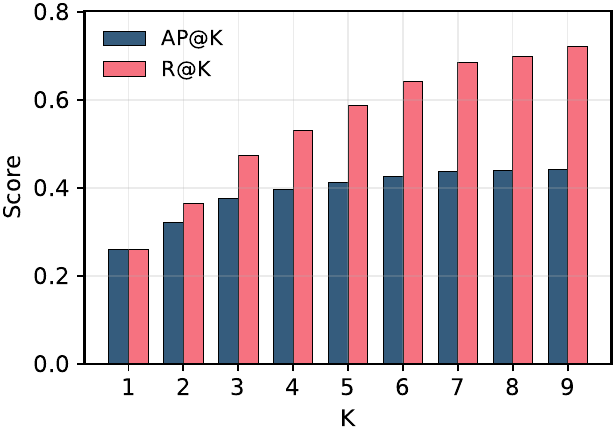}
    \caption{Re-ranking constraints by weighted average similarity}
    \label{fig:rerank_wavg}
\end{figure}

\textbf{Re-ranking constraints by average similarity to retrieved chunks, weighing each similarity by   similarity of chunk to query} $score_{wavg}(c) = \frac{1}{D} \sum_j^D sim(d_j, q) \times sim(c, d_j)$ Building on the rationale of the averaging approach, we recognize that not all documents have the same importance relative to the user query. So we include the similarity of the chunk to the query as a scaling term (which we get for free, from chunk retrieval step). This means that constraints that are similar to chunks that are more relevant to the query get preferential treatment overall. This method shows the best results in Figure \ref{fig:rerank_wavg} improving slightly the normal averaging.

From the five strategies, we can see that in all cases, increasing the number of considered constraints is beneficial, particularly for the recall of the desired constraints. Among them, the best-performing strategy is the one that uses average similarity. As discussed earlier, averaging helps reduce the influence of “outlier” constraints, which explains its superior performance.

\subsubsection{Execution Time}

Due to security-related reasons detailed in Section~\ref{sec:methodology}, redaction is performed post-retrieval of the chunks, and before answering the user query. 
Since the retrieved context has to be redacted separately from the untrusted user query, this operation takes at least 2 LLM calls, depending on the constraints batch size setting. This can be considered impacting compared to the single LLM call required by the baseline with the monolithic prompt.

\begin{figure}[h!]
    \centering

    \begin{subfigure}{0.48\linewidth}
        \centering
        \includegraphics[width=\linewidth]{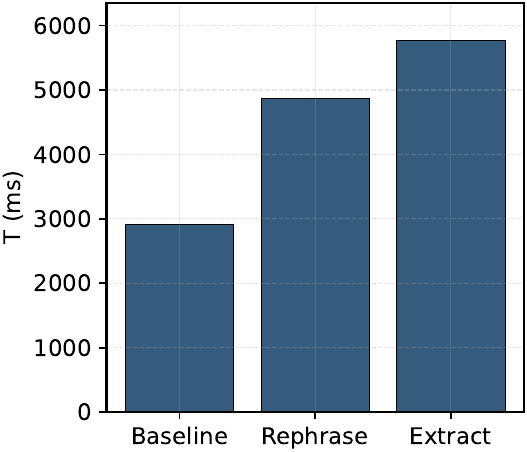}
        \caption{Qwen 2.5 7B q4\_k\_m on Tesla T4}
        \label{fig:times_qwen}
    \end{subfigure}
    \hfill
    \begin{subfigure}{0.48\linewidth}
        \centering
        \includegraphics[width=\linewidth]{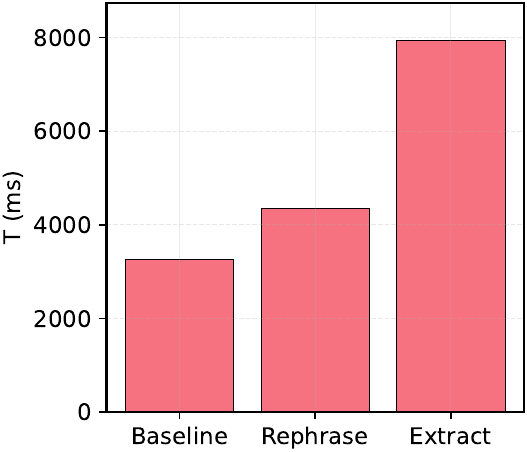}
        \caption{Llama 3 8B q4\_k\_m on Tesla T4}
        \label{fig:times_llama}
    \end{subfigure}

    \caption{Redacting and answering time for different defenses on two models.}
    \label{fig:times_combined}
\end{figure}

To quantify the latency introduced by the pre-redaction, we measure the average time taken to redact and answer questions from the synthetic dataset.
The general trend (as shown in Figure \ref{fig:times_combined}) shows that redacting and answering with the periphrastic technique is faster on average than with the extractive technique. 
This was unexpected, since extracting the sensitive spans (which are usually few) from a text intuitively requires generating less tokens than paraphrasing the full text.
We believe that this is partly due to the usage of a llama-cpp GBNF grammar in our extraction-based redactor to force the model to return the list of sensitive spans in a flat markdown list format.
Applying a grammar to a model's output involves some extra overhead at each token-prediction step, to mask the tokens that do not conform to the grammar's syntax. 
Constraining the model output by a formal grammar is a trade-off: apparently causing an overhead in the general case, but sparing us the complete recomputation of the prompt in cases where the model does not strictly adhere to the formatting instructions.
We do not apply any such grammar to the output of the periphrastic redactor, which simply returns running text without any particular structure. 

Overall, these results are consistent with our earlier robustness analysis: the method that provides the strongest privacy guarantees is also the one that takes the longest to generate a response for the user. This further highlights the trade-off between achieving a high level of privacy and maintaining responsiveness, although the additional time required is not dramatic when compared to the Baseline, even on relatively modest hardware such as a T4 GPU.

\subsubsection{Effect of summarization}

the following experiment is devoted to investigating the effect of hierarchical summarization (similar to what is featured in \cite{sarthi2024raptor}) on privacy preservation. We assume that hierarchical summarization may benefit privacy by two mechanisms:

\begin{itemize}
    \item {Summarization involves generalization of specific (and potentially private) details.}
    \item {Summarization may bring together multiple chunks that would not individually trigger a redaction constraint, but that collectively do.}
\end{itemize}

To conduct this experiment, we considered the entire 20 articles dataset and tested our rephrase approach using Llama-3 8B as a reference model
To investigate the effect of summarization, we implemented the indexing strategy as in \cite{sarthi2024raptor}. Each synthetic article was indexed using a tree with a maximum of three layers.
We experimented with answering the questions, filtering the nodes with three different settings: (i) using only the leaf layer, (ii) using the leaves and the intermediate and top-level summaries, and (iii) using only the summaries (excluding the leaves). We calculated the average privacy and completeness scores for three settings.

From the results in Table~\ref{tab:summarizationResults}, we observe that using summaries leads to only a modest improvement in the privacy score, while causing a comparatively large reduction in completeness.
This indicates that our method remains robust even when we directly include the leaves that contain the original, potentially sensitive content.

\begin{table}[]
\centering
\begin{tabular}{llrr}
\toprule
 & use layers & privacy score & completeness score \\
\midrule
  & [0, 1, 2] & 0.748826 & 0.623125 \\
  & [0] & 0.738732 & 0.637327 \\
  & [1, 2] & 0.813380 & 0.534865 \\
\bottomrule
\end{tabular}
\caption{Performance of the approach using different levels of summarization in the hierarchical tree}
\label{tab:summarizationResults}
\end{table}

\section{Limitations \& Future Work}
\label{sec:limit}

\subsection{Limitations}

We made a basic assumption that the data corpus is not poisoned, (i.e. that the retrieved chunks do not contain any prompt injections). Without this assumption, a malicious retrieved chunk could potentially disable the privacy constraints applied during the pre-redaction step.

Moreover, due to hardware and computational resource constraints, our experimental evaluation was limited to relatively small open-source large language models. The evaluation of the proposed approach on larger-scale and closed-source models would require substantially greater computational resources and, in the case of proprietary systems, access to internal model configurations and inference mechanisms that may not be publicly available. In any case, we argue that, since our approach operates at the retrieval and data-governance layers and is therefore largely model-agnostic, the observed benefits should generalize to larger and closed-source models.

Finally, in this work, we assume that the attacker has no previous background knowledge on the corpus of data. Therefore, we did not conduct a study on de-anonymization attacks that involve multi-turn malicious questioning by an attacker.
For instance, suppose that we are protecting a police record database, and that we have a policy that says: ``hide the names of anyone who committed a misdemeanor (minor offense)".
Let us assume that an attacker knows (either by external means, or by multi-turn interrogation of the RAG pipeline) that there is only one ``John Doe" in the database.
Let us also assume that the attacker is aware that SD-RAG only redacts text post-retrieval.
If the attacker asks the system: ``What misdemeanor did John Doe commit?", the system might answer with: ``A man did not pay for his parking spot". 
But then, given the assumptions on his/her knowledge, the attacker may conclude with a high level of confidence that it really was John Doe who did not pay for his parking spot.

\subsection{Future Work}

A future direction for this work may be to automatically generate and/or iteratively enhance a privacy policy's constraints based on a fuzzier description of the goals of the data owner.

Another direction for improvement is the development of better constraint attachment methods, perhaps computing the similarity scores with a dedicated, fine-tuned cross-encoder (constraint-chunk), rather than relying on off-the-shelf sentence embedding models.

And finally, constraint application could also be improved, either by resorting to fine-tuning a generative LLM on golden examples of text redaction, or by exploring meta-prompting approaches to optimize the extractive and periphrastic redaction prompts.

\section{Conclusion}
\label{sec:conclusion}

In this paper, we have proposed SD-RAG, an approach to Selective Disclosure in Retrieval-Augmented Generation.
The proposed solution extends the related literature by considering security risks derived from prompt injection attacks to the LLM component of the RAG.
Under this attack setting, existing solutions fail to protect against privacy leakage as they typically work to build mechanisms to restrain the generative model from returning the sensitive or access-controlled information.
By contrast, SD-RAG decouples the sanitization and disclosure control of retrieved content from the output generation. To do so, we design a refined data model and retrieval mechanism, which includes a generative-AI based reduction module that enforces security and privacy constraints.
The sanitized content is then provided as input to the LLM, thus obtaining resistance to attack against this last component.
A key innovative contribution of our approach lies in its ability to ingest dynamic security and privacy constraints expressed in human-readable natural language. These constraints are automatically interpreted and encoded within our graph-based data model, where they are semantically linked to the underlying data corpus. This representation enables the enforcement of fine-grained, policy-aware retrieval decisions while maintaining flexibility and accessibility for non-technical users.
Our comprehensive experimental evaluation demonstrates that the proposed approach consistently outperforms existing and related methods, achieving an improvement of up to $58\%$ in the privacy score under the optimal configuration.

\section*{Acknowledgments}

This work was supported by the project ``GoTMaT - Governing Technology to Manage the Transition'' founded by the European Community - Next Generation EU, Mission 4 Component 2 Investment 1.3 - CUP B53C22003990006.

\bibliographystyle{plain}

\end{document}